\documentclass{ieeeaccess}
\usepackage{cite}
\usepackage{amsmath,amssymb,amsfonts}
\usepackage{algorithmic}
\usepackage{graphicx}
\usepackage{textcomp}
\usepackage{url}
\usepackage{xcolor}
\usepackage{booktabs}
\usepackage{multirow}
\usepackage{pifont}
\usepackage{colortbl}
\newcommand{\cmark}{\ding{51}}
\newcommand{\pmark}{(\ding{51})}
\newcommand{\xmark}{\textcolor{black!30}{\ding{55}}}
\DeclareUnicodeCharacter{2212}{-}

\newcommand{\thd}[1]{\multicolumn{1}{c}{\begin{tabular}[b]{@{}c@{}}#1\end{tabular}}}

\usepackage[shortcuts,acronym]{glossaries} 
\usepackage[shortcuts,acronym]{glossaries}

\newacronym{uwb}{UWB}{ultra-wideband}
\newacronym{ir-uwb}{IR-UWB}{impulse-radio ultra-wideband}
\newacronym{fmcw}{FMCW}{frequency-modulated continuous wave}
\newacronym{ble}{BLE}{bluetooth low energy}
\newacronym{mmwave}{mmWave}{millimeterwave}
\newacronym{gnss}{GNSS}{global navigation satellite system}
\newacronym{elu}{ELU}{Exponential Linear Unit}
\newacronym{fr3}{FR3}{frequency range 3}
\newacronym{ch5}{CH5}{channel 5}
\newacronym{ch9}{CH9}{channel 9}
\newacronym{RDM}{RDM}{Range Doppler Map}
\newacronym{jcas}{JCAS}{joint communication and sensing}
\newacronym{isac}{ISAC}{integrated sensing and communication}

\newacronym{ioe}{IoE}{Internet of Everything}
\newacronym{iot}{IoT}{Internet of Things}
\newacronym{iiot}{IIoT}{Industrial Internet of Things}

\newacronym{psd}{PSD}{Power Spectral Density}

\newacronym{afsiw}{AFSIW}{Air-Filled Substrate-Integrated-Waveguide}
\newacronym{cbs}{CBS}{Cavity-Backed Slot}
\newacronym{dra}{DRA}{Dielectric Resonator Antenna}
\newacronym{mimo}{MIMO}{Multiple Input Multiple Output}
\newacronym{hpbw}{HPBW}{Half-Power Beamwidth}
\newacronym{ftbr}{FTBR}{Front-to-Back Ratio}
\newacronym{sff}{SFF}{System Fidelity Factor}
\newacronym{dee}{DEE}{Distance Estimation Error}
\newacronym{fov}{FOV}{field-of-view}
\newacronym{qm}{QM}{Quarter-Mode}

\newacronym{rtls}{RTLS}{real-time location system}
\newacronym{tof}{ToF}{Time of Flight}
\newacronym{rtt}{RTT}{Round Trip Time}
\newacronym{toa}{ToA}{Time of Arrival}
\newacronym{twr}{TWR}{Two-way Ranging}
\newacronym{tdoa}{TDoA}{Time Difference of Arrival}
\newacronym{pdoa}{PDoA}{Phase Difference of Arrival}
\newacronym{aoa}{AoA}{Angle of Arrival}
\newacronym{acir}{ACIR}{accumulated channel impulse response}
\newacronym{cir}{CIR}{channel impulse response}
\newacronym{csi}{CSI}{channel state information}
\newacronym{dop}{DOP}{dilution of precision}

\newacronym{mae}{MAE}{mean absolute error}
\newacronym{rmse}{RMSE}{root mean square Error}

\newacronym{music}{MUSIC}{Multiple Signal Classification}
\newacronym{fft}{FFT}{fast fourier transform}
\newacronym{stft}{STFT}{short-time fourier transform}
\newacronym{ekf}{EKF}{Extended Kalman Filter}
\newacronym{iq}{IQ}{in-phase and quadrature}
\newacronym{snr}{SNR}{signal to noise}

\newacronym{rtm}{RTM}{range-time map}
\newacronym{rdm}{RDM}{range-Doppler map}
\newacronym{tdm}{TDM}{time-Doppler map}

\newacronym{cpi}{CPI}{coherent processing interval}

\newacronym{mti}{MTI}{Moving Target Indication}

\newacronym{cpu}{CPU}{central processing unit}
\newacronym{gpu}{GPU}{graphics processing unit}
\newacronym{fpu}{FPU}{floating-point unit}
\newacronym{dat}{DAT}{dual-antenna tile}
\newacronym{cots}{COTS}{commercial off-the-shelf}
\newacronym{trx}{TRX}{transceiver}
\newacronym{ic}{IC}{integrated circuit}
\newacronym{pcb}{PCB}{Printed Circuit Board}
\newacronym{tcxo}{TCXO}{temperature-controlled crystal oscillator}
\newacronym{pu}{PU}{processing unit}
\newacronym{pmic}{PMIC}{power management integrated circuit}

\newacronym{spi}{SPI}{Serial Peripheral Interface}
\newacronym{i2c}{I2C}{Inter-Integrated Circuit}
\newacronym{csn}{CSn}{Chip Select}
\newacronym{sda}{SDA}{Serial Data}
\newacronym{scl}{SCL}{Serial Clock}
\newacronym{gpio}{GPIO}{general-purpose input-output}
\newacronym{io}{IO}{input-output}

\newacronym{rtos}{RTOS}{real-time operating system}
\newacronym{soc}{SoC}{system-on-chip}
\newacronym{rf}{RF}{radio frequency}
\newacronym{gcpw}{GCPW}{grounded coplanar waveguide}
\newacronym{tx}{TX}{transmit}
\newacronym{rx}{RX}{receive}

\newacronym{los}{LOS}{line-of-sight}
\newacronym{nlos}{NLOS}{non-line-of-sight}

\newacronym{roi}{ROI}{region of interest}

\newacronym{dk}{DK}{development kit}
\newacronym{sram}{SRAM}{static random access memory}

\newacronym{fpga}{FPGA}{field-programmable gate array}

\newacronym{sdr}{SDR}{Software Defined Radio}

\newacronym{fp}{FP}{first path}

\newacronym{har}{HAR}{human activity recognition}
\newacronym{ml}{ML}{machine learning}
\newacronym{ai}{AI}{artificial intelligence}
\newacronym{cnn}{CNN}{convolutional neural network}
\newacronym{lstm}{LSTM}{long short-term memory}
\newacronym{dnn}{DNN}{deep neural network}
\newacronym{nn}{NN}{neural network}

\newacronym{rr}{RR}{respiration rate}
\newacronym{br}{BR}{breathing rate}
\newacronym{hr}{HR}{heart rate}
\newacronym{rbm}{RBM}{random body movement}
\newacronym{bpm}{BPM}{breaths per minute}

\newacronym{pll}{PLL}{phase locked loop}
\newacronym{shr}{SHR}{synchronization header}
\newacronym{phr}{PHR}{physical layer header}
\newacronym{psdu}{PSDU}{physical layer service data unit}

\newacronym{mac}{MAC}{medium access control}
\newacronym{rs}{RS}{reed solomon}

\newacronym{tts}{TTS}{text-to-speech}

\newacronym{lopo}{LOPO}{leave-one-person-out}
\newacronym{loso}{LOSO}{leave-one-scenario-out}
\newacronym{lo2so}{LO2SO}{leave-two-scenarios-out}
\newacronym{lobpo}{LOBPO}{leave-one-bed-position-out}
\newacronym{loro}{LORO}{leave-one-room-out}

\newacronym{macc}{MAC}{multiply-accumulate operation}

\newacronym{ofdm}{OFDM}{Orthogonal Frequency Division Multiplexing}
\usepackage{cleveref} 
\usepackage{soul} 
\crefname{figure}{Fig.}{Figs.}
\Crefname{figure}{Fig.}{Figs.}
\crefname{table}{Table}{Tables}
\Crefname{table}{Table}{Tables}
\crefname{section}{Section}{Sections}
\Crefname{section}{Section}{Sections}

\usepackage{silence}
\DeclareTextFontCommand{\textbf}{\bfseries}

\usepackage{bm}
\makeatletter
\AtBeginDocument{\DeclareMathVersion{bold}
\SetSymbolFont{operators}{bold}{T1}{times}{b}{n}
\SetSymbolFont{NewLetters}{bold}{T1}{times}{b}{it}
\SetMathAlphabet{\mathrm}{bold}{T1}{times}{b}{n}
\SetMathAlphabet{\mathit}{bold}{T1}{times}{b}{it}
\SetMathAlphabet{\mathbf}{bold}{T1}{times}{b}{n}
\SetMathAlphabet{\mathtt}{bold}{OT1}{pcr}{b}{n}
\SetSymbolFont{symbols}{bold}{OMS}{cmsy}{b}{n}
\renewcommand\boldmath{\@nomath\boldmath\mathversion{bold}}}
\makeatother

\def\BibTeX{{\rm B\kern-.05em{\sc i\kern-.025em b}\kern-.08em
    T\kern-.1667em\lower.7ex\hbox{E}\kern-.125emX}}

\providecommand{\dset}[3]{\makebox[1em][r]{#1}\ /\ \makebox[1em][r]{#2}\ /\ \makebox[0.5em][l]{#3}}

\begin{document}
\history{Date of publication xxxx 00, 0000, date of current version xxxx 00, 0000.}
\doi{10.1109/ACCESS.2024.0429000}

\title{One UWB Radar, Many Views: Comparing and Fusing Signal Representations for Human Activity Recognition}
\author{\uppercase{Jelle De Moerloose}\authorrefmark{1,*}, \uppercase{Anton Lambrecht}\authorrefmark{1,*},
\uppercase{Adnan Shahid}\authorrefmark{1}, \IEEEmembership{Senior Member, IEEE}, AND \uppercase{Eli De Poorter}\authorrefmark{1}}

\address[1]{IDLab, Department of Information Technology, Ghent University--imec, Ghent, Belgium (e-mail: jelle.demoerloose@ugent.be)}

\tfootnote{The research that led to these results has received funding from the DistriMuSe project (HORIZON-KDT-JU-2023-2-RIA) with Grant No 101139769, the imec ICON project NG-UWB (Agentschap Innoveren en Ondernemen project nr. HBC.2024.0882) and the NAV-ALERT project from the Belgian Defense through contract number 24DEFRA009.\\ $^{*}$Jelle De Moerloose and Anton Lambrecht contributed equally to this work.}

\markboth
{De Moerloose \headeretal: One UWB Radar, Many Views}
{De Moerloose \headeretal: One UWB Radar, Many Views}

\corresp{Corresponding author: Jelle De Moerloose (e-mail: jelle.demoerloose@ugent.be).}

\begin{abstract}

IEEE 802.15.4z-compliant impulse-radio ultra-wideband (IR-UWB) transceivers are increasingly built into consumer devices, enabling contactless, privacy-preserving human activity recognition (HAR) in healthcare. Before activity classification, the signal is typically converted into range-time, range-Doppler, or time-Doppler maps, each exposing different spatial and motion cues. Existing comparisons often apply shared processing across representations, leaving it unclear whether their rankings hold after independent tuning. It is also unknown whether rankings and fusion gains generalize across environments and activities, and at what computational cost. Consequently, practitioners lack guidance on which representation and processing suit a given HAR deployment. We release an open dataset of raw IR-UWB measurements collected by a ceiling-mounted radar from 44 participants during scripted activity sequences in 12 layouts across two bedrooms. Using a statistically guided, staged search, we independently tune each representation across five processing and deployment settings. We then compare individual and fused performance under layout and room shifts and quantify computational cost. Across over 1{,}000 cross-validated models, independent tuning proves essential, as optimal processing differs per representation and processing choices shift macro-F1 by up to 26 points, four times the gap between representations. Once tuned, the range-Doppler map leads for unseen layouts (85.1\% macro-F1), while time-Doppler map does for unseen rooms (72.7\%). Fusing all three maps raises these scores to 89.1\% and 80.0\%, respectively. Lower sampling rates and narrower Doppler bands reduce latency by 38--51\% with at most a 1.5-point macro-F1 loss. The best representation for UWB HAR is therefore a deployment decision, not a fixed property.
\end{abstract}


\titlepgskip=-21pt

\maketitle

\section{Introduction}
\label{sec:intro}

\PARstart{H}{uman} activity recognition (\acs{har})\glsunset{har} systems help caregivers in care facilities understand daily routines, identify changes in mobility and functional ability, and respond more quickly to potential emergencies, while also supporting independent living at home \cite{radar_survey}. However, camera-based monitoring raises privacy concerns by capturing identifiable images, while wearable sensors depend on patient compliance, which can be particularly low among elderly or cognitively impaired individuals \cite{huang2025}. \Ac{rf} sensing uses reflected radio waves to detect movement without requiring cameras or on-body devices, offering an unobtrusive and privacy-friendly approach to HAR \cite{radar_survey}.

The choice of \ac{rf} technology involves trade-offs between sensing performance, deployment practicality and cost. For example, a \Ac{fmcw} radar is well established for sensing, while Wi-Fi can leverage existing communication infrastructure for sensing but is limited by its narrower bandwidth. \Ac{ir-uwb} provides finer range resolution, while commercial IEEE 802.15.4z-compliant transceivers additionally support standardized ranging and communication~\cite{coppens_overview_2022, lambrecht2026comparison}. This combination makes \ac{ir-uwb} particularly attractive for integrating sensing and communication on a single commercial platform, potentially reducing deployment costs.

Radar-based \ac{har} classifies radar data into daily activities, such as walking or getting out of bed, typically using deep learning~\cite{radar_survey}. Most work has focused on improving model architectures~\cite{swinT}, while the input representation and its processing have received less attention. This is a critical gap, as no model architecture can recover information discarded during preprocessing. Within \ac{ir-uwb}, the \acp{cir} within an observation window of a few seconds are typically converted into a \ac{rtm}, \ac{rdm}, or \ac{tdm} before classification~\cite{complex_CNN}. These representations balance spatial and motion information differently, resulting in representation-specific optimal processing settings (e.g., clutter removal) and deployment settings (e.g., slow-time sampling rate and carrier frequency). Existing comparisons~\cite{swinT, complex_CNN, ding2021, Ayaz2025}, however, apply fixed settings across representations, which can favor one representation over another. They also do not distinguish between deployment scenarios, such as a changed furniture layout or an unseen room, although the best representation may differ between them. Moreover, since the representations emphasize different information, they may be complementary. Fusion has been explored for radar \ac{har}~\cite{classic, signal_adapted_CNN,DL-RDM}, but which representation benefits which activity, and whether this complementarity holds when generalizing to unseen rooms or room layouts, has not been systematically analyzed. As a result, there is limited guidance on which representation and settings to use for a given deployment. To address this, this paper makes the following contributions:
\begin{itemize}
    \item A systematic evaluation of \ac{rtm}, \ac{rdm}, and \ac{tdm}
    representations for \ac{ir-uwb} \ac{har}, comparing their performance
    under realistic deployment changes (unseen furniture layouts and new
    rooms) and revealing their representation-dependent sensitivity to
    key processing and deployment settings: clutter removal, \ac{fft}
    windowing, slow-time sampling rate, Doppler bandwidth, and carrier
    frequency.
    \item An investigation of cross-representation fusion, uncovering complementary activity-specific information across \ac{rtm}, \ac{rdm}, and \ac{tdm} and quantifying its contribution to recognition performance.
    \item A comprehensive analysis of accuracy, latency, and model complexity, identifying configurations that reduce processing latency by 38–51\% with at most a 1.5-point drop in macro-F1.
    \item An open-access \ac{ir-uwb} dataset from 44 participants across 12 layouts in two rooms, extending our earlier single-room dataset~\cite{lambrecht2026comparison} with a second room and channel~9 recordings, supporting reproducible benchmarking and future research on \ac{ir-uwb}-based \ac{har}.\footnote{\url{https://github.com/anton-lambrecht/IDLab_iWiNe-Activity-Recognition-Datasets}}
\end{itemize}

\section{Related Work}
\label{sec:related}

\begin{table*}[t]
\centering
\footnotesize
\renewcommand{\arraystretch}{1.15}
\setlength{\tabcolsep}{3.3pt}
\caption{Comparison of the IR-UWB \ac{har} studies closest to this work. Prior studies generally use fixed processing and deployment conditions. This work evaluates all three representations and their fusion on IEEE 802.15.4z \ac{cots} hardware, across processing choices, computational costs, and layout and room shifts.}
\label{tab:related_work_scope}
\begin{tabular}{@{}l l c c c cccc ccccc cc l@{}}
\toprule
& & & &
& \multicolumn{4}{c}{Representation}
& \multicolumn{5}{c}{Processing factor varied}
& \multicolumn{2}{c}{Cost per repr.}
& \\
\cmidrule(lr){6-9}\cmidrule(lr){10-14}\cmidrule(lr){15-16}
Ref. & \multicolumn{1}{l}{\begin{tabular}[b]{@{}l@{}}UWB\\radar\end{tabular}} & \thd{BW\\(GHz)} & \thd{802.15.4\\COTS} & \thd{Subj./\\layouts/\\rooms}
& RTM & RDM & TDM & Fusion
& \thd{RF\\ch.} & \thd{Clutter\\rem.} & \thd{FFT\\window} & \thd{Samp.\\rate} & \thd{Dopp.\\band}
& \thd{Pre-\\proc.} & Model
& Generalization \\
\midrule
\cite{classic}              & X4M03  & 2.5 & \xmark & \dset{3}{1}{1}   & \xmark & \xmark & \xmark & \cmark & \xmark & \xmark & \xmark & \xmark & \xmark & \xmark & \xmark & Random \\
\cite{complex_CNN}          & P410   & 2.2 & \xmark & \dset{15}{1}{1}  & \cmark & \cmark & \cmark & \xmark & \xmark & \xmark & \xmark & \xmark & \xmark & \xmark & \xmark & Subj. \\
\cite{signal_adapted_CNN}   & XeThru & 1.4 & \xmark & \dset{30}{7}{7}  & \cmark & \cmark & \xmark & \cmark & \xmark & \xmark & \xmark & \xmark & \xmark & \xmark & \xmark & Subj., room, height \\
\cite{swinT}                & P440   & 1.7 & \xmark & \dset{22}{3}{3}  & \cmark & \cmark & \cmark & \xmark & \xmark & \xmark & \xmark & \xmark & \xmark & \cmark & \xmark & Subj., room \\
\cite{cir_concatenation}    & DW1000 & 0.5 & \cmark & \dset{4}{3}{1} & \cmark & \xmark & \xmark & \xmark & \xmark & \xmark & \xmark & \xmark & \xmark & \xmark & \xmark & Random \\
\cite{time_freq_comparison} & P410   & 2.2 & \xmark & \dset{15}{1}{1}  & \xmark & \xmark & \cmark & \xmark & \xmark & \xmark & \xmark & \xmark & \xmark & \xmark & \cmark & Subj. \\
\cite{DL-RDM}               & P410   & 2.2 & \xmark & \dset{14}{2}{2}  & \pmark & \cmark & \cmark & \cmark & \xmark & \xmark & \xmark & \xmark & \xmark & \xmark & \xmark & Subj. \\
\cite{lambrecht2026comparison} & DW3000 & 0.5 & \cmark & \dset{20}{6}{1} & \xmark & \cmark & \xmark & \xmark & \xmark & \xmark & \xmark & \xmark & \xmark & \xmark & \xmark & Subj., layout \\
\midrule
\rowcolor{black!6}
\textbf{Ours}               & \textbf{DW3000} & 0.5 & \cmark & \dset{44}{12}{2} & \cmark & \cmark & \cmark & \cmark & \cmark & \cmark & \cmark & \cmark & \cmark & \cmark & \cmark & Layout, room \\ 
\bottomrule
\end{tabular}
 
\vspace{2pt}
\end{table*}

The differences and complementary nature between \acp{rtm}, \acp{rdm}, and \acp{tdm} have been investigated in the context of \ac{fmcw} radar. Ding et al.~\cite{ding2021} fuse all
three representations using learnable weights and show through ablation studies that no
single representation consistently performs best across all activities. Ayaz
et al.~\cite{Ayaz2025} compare an \ac{rtm} and two \acp{tdm} across four \ac{cnn} backbones
and, unlike most prior work, also report the preprocessing cost of each. However, both
studies evaluate the representations using only a single set of preprocessing parameters.
Both also split samples randomly, so the same subjects and rooms appear in training and test
sets. As a result, there is little guidance on which representation suits a given deployment. Moreover, \ac{fmcw} radars are designed primarily for sensing, whereas IEEE 802.15.4z-compliant \ac{ir-uwb} modules are designed for ranging and communication~\cite{lambrecht2026comparison}. It is therefore unclear whether conclusions drawn for \ac{fmcw} radar carry over to such modules.

\Cref{tab:related_work_scope} summarizes the \ac{ir-uwb} studies closest to this work. 
In these studies, the signal representation is typically treated as a design choice rather than an object of study. Some evaluate a single input: an \ac{rtm}~\cite{cir_concatenation}, an \ac{rdm}~\cite{lambrecht2026comparison}, or a fused set of hand-crafted \ac{rtm} and \ac{tdm} features~\cite{classic}. Swarubini et al.~\cite{time_freq_comparison} compare four time-frequency estimators but stay within the micro-Doppler domain. Chen et al.~\cite{signal_adapted_CNN} and Wang et al.~\cite{DL-RDM} combine two representations in a fusion model without examining when either is preferable. Yang et al.~\cite{complex_CNN} and Li et al.~\cite{swinT} compare all three as standalone inputs, and both find the \ac{rtm} best.

These comparisons share a key limitation: each is made at a single processing operating point. Li et al.~\cite{swinT} compare two ways of computing each representation, such as restricting the \ac{tdm} to detected target ranges or adding a time axis to the \ac{rdm}. Beyond such variants, however, the processing is fixed in every study of \Cref{tab:related_work_scope}~\cite{classic,complex_CNN,signal_adapted_CNN,swinT,cir_concatenation,time_freq_comparison,DL-RDM,lambrecht2026comparison}, as it is in the \ac{fmcw} studies. None varies the carrier frequency, clutter removal, \ac{fft} window, slow-time sampling rate, or Doppler band. Yet these factors affect each representation differently. Static clutter spreads across the entire \ac{rtm}, whereas in \acp{rdm} and \acp{tdm} it concentrates near zero Doppler. The \ac{fft} window affects only \acp{rdm} and \acp{tdm}, as the \ac{rtm} involves no Fourier transform. The slow-time sampling rate limits the Doppler range of \acp{rdm} and \acp{tdm}, but only reduces the temporal resolution of the \ac{rtm}. A setting that suits one representation may therefore disadvantage another.

On resource-constrained devices, computational cost can weigh as heavily as accuracy in the choice of representation. However, existing studies rarely consider it, and those that do report only part of it: Li et al.~\cite{swinT} measure the computation time of individual representations, while Swarubini et al.~\cite{time_freq_comparison} report model inference time. A comparison of end-to-end computational cost, covering the full processing pipeline and the model, is still missing.

Generalization to unseen environments remains less explored than generalization to unseen subjects, despite being essential for deploying \ac{har} in new rooms and layouts without costly environment-specific data collection and retraining. Only Chen et al.~\cite{signal_adapted_CNN} and Li et al.~\cite{swinT} evaluate unseen-room performance, without comparing representations across environments. The importance of this gap is highlighted by El Hail et al.~\cite{el_hail_radar-based_2025}, who report a 30--39-point F1 drop between rooms for an \ac{fmcw}-based model. Although they mitigate this through data augmentation and domain adaptation, their representation remains fixed to the \ac{tdm}. Which representation best generalizes across environments therefore remains unknown.

Few studies adopt a deployment-ready setup. Besides our earlier work~\cite{lambrecht2026comparison}, only Wang and Basiri~\cite{cir_concatenation} use IEEE 802.15.4-compliant \ac{cots} modules, whereas the others rely on dedicated sensing radars with considerably more bandwidth. Likewise, only our earlier work and Chen et al.~\cite{signal_adapted_CNN} use ceiling-mounted radars, facilitating non-intrusive healthcare integration at the cost of a smaller radar cross-section~\cite{lambrecht2026comparison}. Additionally, the datasets in \Cref{tab:related_work_scope} include at most 30 subjects, and five of the eight cover a single room.

To address these gaps, we compare \acp{rtm}, \acp{rdm}, and \acp{tdm}, individually and through feature-level fusion,using a single ceiling-mounted IEEE 802.15.4z-compliant \ac{cots} platform, controlling for hardware differences across representations. As summarized in \Cref{tab:related_work_scope}, we investigate five processing and deployment factors, tuning the settings independently for each representation. We evaluate generalization to unseen layouts and rooms, quantify preprocessing time, inference time, and model complexity, and use statistical tests to assess performance differences.

\section{Signal Model, Representations, and Preprocessing}
\label{sec:sysrem}

This section introduces the radar signal model and the representations derived from it. We describe the complex \ac{cir} and its fast- and slow-time structure, the resulting range-time, range-Doppler, and time-Doppler representations, and the processing and deployment settings varied in our experiments.

\subsection{Radar Signal Model}
\label{sec:signal_model}
\begin{figure*}[!t]
  \centering
  \includegraphics[width=\textwidth]{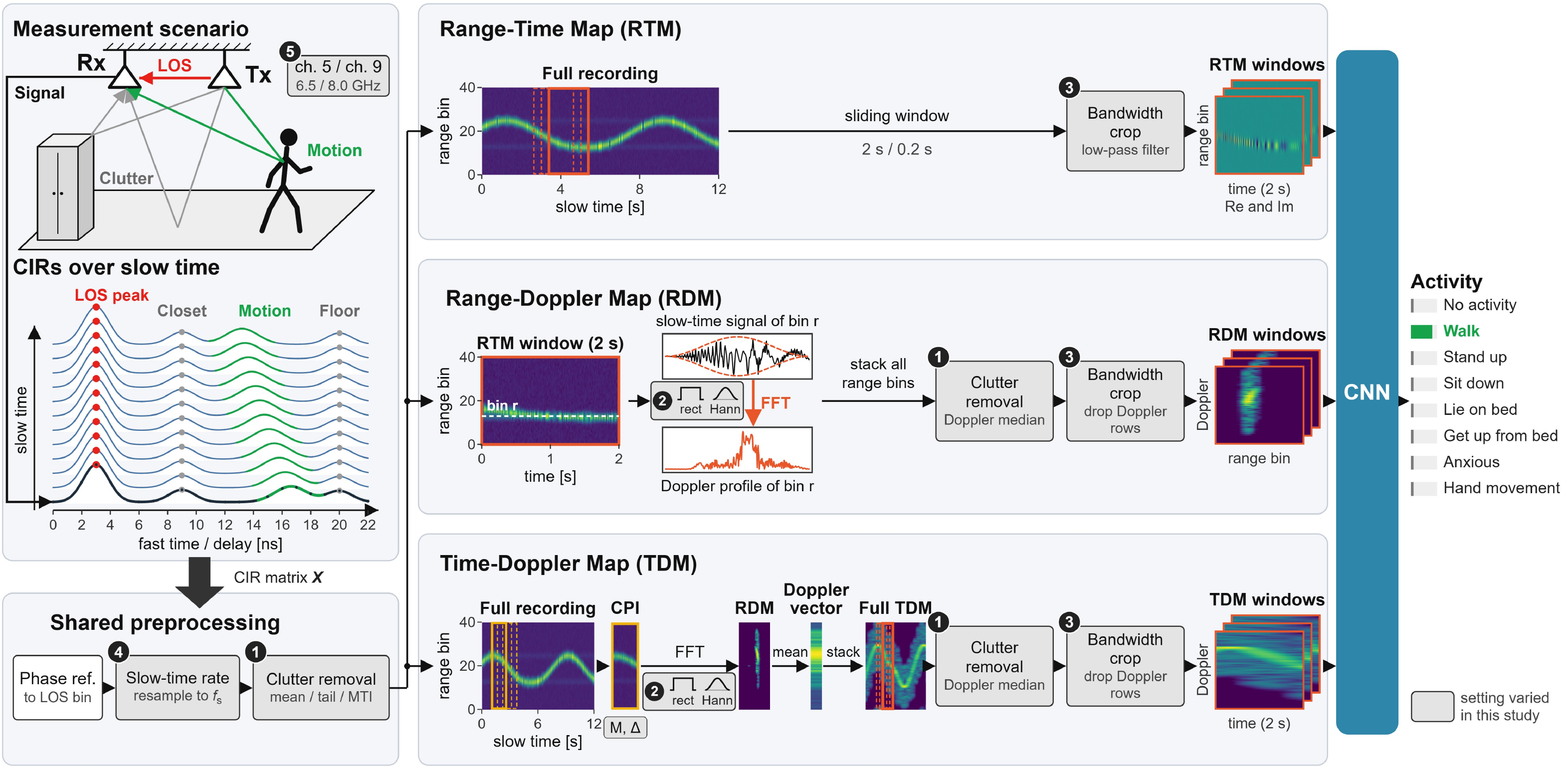}
    \caption{Signal model and representation pipeline. With a ceiling-mounted transmitter and receiver, each transmission produces a \ac{cir} containing \ac{los}, static-clutter, and motion reflections. Stacking \acp{cir} over slow time produces an \ac{rtm}. A slow-time \ac{fft} per range bin yields an \ac{rdm}, while \acp{fft} over short \acp{cpi}, averaged across range and stacked over time, form a \ac{tdm}. Each representation is windowed and classified by a separately trained \ac{cnn} with the same architecture. Grey blocks mark the preprocessing and deployment settings varied in this study, numbered as in \cref{sec:preprocessing}. }
  \label{fig:cir_stack}
\end{figure*}

An \ac{ir-uwb} radar measures the response of the environment to short pulses. The transmitter and receiver are mounted close together on the ceiling, forming a pseudo-monostatic geometry, as shown in \cref{fig:cir_stack}. The receiver captures the direct \ac{los} path and the echoes from scatterers in the scene, each delayed in proportion to its path length. In IEEE 802.15.4z, each packet carries a preamble of many coded pulses, which the receiver correlates and accumulates into one \ac{cir}. Sampled along the delay axis, the fast time, it is given by
\begin{equation}
    h[k] = e^{j\theta} \sum_{p=0}^{P-1} \alpha_p \, e^{j\phi_p} \, \delta[k - k_p] + w[k],
    \label{eq:cir}
\end{equation}
where $\alpha_p$, $\phi_p$, and $k_p$ are the amplitude, phase, and fast-time delay bin
of the $p$-th path, with $p=0$ the \ac{los} path, and $w[k]$ is additive noise. The term
$\theta$ is a phase offset common to all bins of one \ac{cir}, caused by the carrier
frequency offset and clock drift between the unsynchronized transmitter and receiver. It changes from packet to packet and would corrupt the slow-time phase progression that the following representations rely on. Since the \ac{los} path is static,
referencing every bin to its phase removes $\theta$~\cite{demoerloose2026mmlevelaccurateuwbradar},
\begin{equation}
    \hat{h}[k] = h[k]\, e^{-j \angle h[k_{\mathrm{LOS}}]}.
    \label{eq:phase_ref}
\end{equation}
Packets are transmitted at a slow-time interval $T_s = 1/f_s$, where $f_s$ is the slow-time sampling rate. Stacking $N_s$
phase-referenced \acp{cir} gives
\begin{equation}
    \mathbf{X} \in \mathbb{C}^{N_s \times N_f},
    \qquad
    [\mathbf{X}]_{m,k} = \hat{h}_m[k],
    \label{eq:cir_matrix}
\end{equation}
whose rows index slow time and whose $N_f$ columns index fast-time bins, mapping
to range. Each entry decomposes as
\begin{equation}
    [\mathbf{X}]_{m,k} = s_k + d_{m,k} + g_{m,k} + \nu_{m,k},
    \label{eq:decomposition}
\end{equation}
where $s_k$ is the time-invariant contribution of static reflectors, such as the closet and floor in \cref{fig:cir_stack}, $d_{m,k}$
the variation caused by human motion, $g_{m,k}$ the contribution of other moving
scatterers such as fans or swaying curtains, and $\nu_{m,k}$ thermal noise. Only $s_k$ is constant along slow time, while the other terms vary along both
slow and fast time. Without phase referencing, $s_k$ would not be time-invariant.

\subsection{Signal Representations}
\label{sec:signal_rep}

For classification, the \ac{cir} matrix is divided into overlapping windows of $W$ consecutive \acp{cir}, each spanning $T_{\mathrm{obs}} = W/f_s$ and advancing by a fixed stride. Each window is converted into an \ac{rtm}, \ac{rdm}, or \ac{tdm} that serves as one \ac{cnn} input, as shown in \cref{fig:cir_stack}. Window and \acf{cpi} lengths are listed in \cref{sec:methodology}.

\subsubsection{\Acf{rtm}}
\label{subsec:rtm}
The \ac{rtm} keeps a window of $\mathbf{X}$ without further transformation. It shows how the reflected signal at each range evolves over time, so a moving person appears as a trace that shifts and fluctuates across range bins. It preserves full range and time resolution but leaves motion implicit in the slow-time evolution, which the network must learn. Most prior work keeps only the magnitude~\cite{signal_adapted_CNN,classic,cir_concatenation}. We use the real and imaginary parts, since the corrected phase can track displacement within
a fraction of a wavelength~\cite{demoerloose2026mmlevelaccurateuwbradar}.

\subsubsection{\Acf{rdm}}
\label{subsec:rdm}
The \ac{rdm} summarizes the motion within a window by its Doppler frequency at
each range, rather than by how the signal changes over time. For each range bin, an \ac{fft} is applied along the slow-time axis,
\begin{equation}
    D[k,l] = \left| \sum_{m=0}^{W-1} a[m]\, [\mathbf{X}]_{m,k}\,
    e^{-j 2\pi m l / W} \right|,
    \label{eq:rdm}
\end{equation}
where $l$ indexes Doppler frequency bins and $a[m]$ is a window function, so $D[k,l]$ is the intensity at range bin $k$ and Doppler frequency bin $l$. Radial velocity $v_r$ toward the radar yields $f_D \approx 2 v_r/\lambda$, with $\lambda$ the carrier wavelength, placing approaching targets at positive Doppler, receding targets at negative Doppler, and static returns at zero Doppler. The Doppler resolution is $f_s/W$ and the axis spans $\pm f_s/2$, so faster motion aliases into this band. A longer window provides finer velocity resolution, but averages motion over a longer
interval, reducing the temporal resolution. Only the magnitude is used, since the phase
of each coefficient depends on the window start and differs between windows for
identical motion.

\subsubsection{\Acf{tdm}}
\label{subsec:tdm}
The \ac{tdm} is a micro-Doppler spectrogram, a widely used representation in radar-based activity recognition~\cite{ding2021,Ayaz2025,time_freq_comparison}. It shows how the Doppler content of the scene evolves over time. To construct it, the slow-time samples in each observation window are divided into overlapping \acp{cpi}, each containing \(M < W\) consecutive \acp{cir} (\cref{fig:cir_stack}). For each \ac{cpi}, a short-time \ac{rdm} is formed by applying the slow-time \ac{fft} independently at each range bin. Averaging the magnitude spectra across range produces one Doppler vector per \ac{cpi}. The \ac{cpi} advances through the observation window with stride \(\Delta < M\). Stacking these vectors in temporal order forms the \ac{tdm}. Thus, compared with the \ac{rdm}, the \ac{tdm} trades range localization for a time-resolved view of Doppler content.

\subsection{Processing and Deployment Settings}
\label{sec:preprocessing}
This subsection defines the processing and deployment settings varied in our experiments, numbered (1)--(5) in \cref{fig:cir_stack}. The swept values are listed in \cref{sec:methodology}.

\subsubsection{Clutter Removal}
Clutter removal suppresses the static term $s_k$ in \cref{eq:decomposition} by
subtracting an estimate $\hat{s}_k$ from every \ac{cir},
\begin{equation}
    \hat{s}_k = \frac{1}{|\mathcal{R}|} \sum_{m \in \mathcal{R}} [\mathbf{X}]_{m,k},
    \label{eq:static_est}
\end{equation}
where $\mathcal{R}$ is a set of reference \acp{cir}. Mean subtraction uses the current window for $\mathcal{R}$, while tail subtraction uses
$T_{\mathrm{ref}}=1$\,s of empty-room \acp{cir} from each recording's tail. Tail subtraction requires an empty-room reference, which can be acquired at installation and refreshed after layout changes, for example when presence detection indicates an empty room, whereas the other estimators rely solely on the incoming signal. The \ac{mti} filter subtracts consecutive
pulses, $[\mathbf{X}]_{m,k} - [\mathbf{X}]_{m-1,k}$, removing $s_k$ but also attenuating slow motion. 

For the \ac{rdm} and \ac{tdm}, Doppler-median subtraction can be applied after the
\ac{fft}. Let $D_t$ denote either the \ac{rdm} of the window starting at time $t$, computed with \cref{eq:rdm}, or the \ac{tdm} column of the \ac{cpi} starting
at $t$. Each map is reduced by the median of all maps within
a centered interval of $T_{\mathrm{med}} = 10$\,s,
\begin{equation}
    \tilde{D}_t = \max\!\Big(0,\; D_t - \operatorname*{median}_{|t' - t| \le T_{\mathrm{med}}/2} D_{t'}\Big),
    \label{eq:doppler_median}
\end{equation}
where the median and the maximum act element-wise. This long horizon median also removes persistent motion such as a rotating fan. The centered interval introduces a
latency of $T_{\mathrm{med}}/2$, which a causal interval over the preceding
$T_{\mathrm{med}}$ avoids and is expected to perform similarly.

\subsubsection{Window Function}
We compare two choices for the window $a[m]$ in \cref{eq:rdm}, rectangular and Hann. Hann is commonly used in prior work~\cite{swinT,time_freq_comparison,Ayaz2025}. It suppresses spectral leakage but widens the main lobe, trading Doppler resolution for a cleaner spectrum.

\subsubsection{Doppler Bandwidth Cropping}
For the \ac{rdm} and \ac{tdm}, Doppler bandwidth is reduced by removing rows from the Doppler maps corresponding to Doppler components outside the selected frequency range. This directly reduces the input size without additional computational cost. The \ac{rtm} has no Doppler axis, so the same band limit requires a fourth-order zero-phase Butterworth low-pass filter along slow time, which adds computation without reducing the input size.

\subsubsection{Slow-time Sampling Rate}
Lower slow-time sampling rates are emulated by linearly interpolating $\mathbf{X}$ along
slow time onto a coarser uniform grid. A lower rate reduces spectrum usage, the amount of collected data and
computational complexity but narrows the Doppler axis to $\pm f_s/2$. No anti-aliasing filter
is applied, as a radar at a lower slow-time sampling rate also samples without
one, so in both cases Doppler content beyond $\pm f_s/2$ aliases into the band.

\subsubsection{Carrier Frequency}
The DW3000 supports channel~5 (6.5\,GHz) and channel~9 (8.0\,GHz), both with 499.2\,MHz bandwidth, so range resolution is identical. The shorter wavelength of channel~9 increases the phase change and Doppler shift of the same motion by 23\%. This improves sensitivity to small motion and velocity resolution, but causes earlier aliasing at a given $f_s$ and higher propagation loss.

\section{Dataset and Hardware}
\label{sec:dataset_hardware}

This section describes the data collection campaign and the radar
hardware used throughout the paper.

\subsection{Dataset}
\label{sec:dataset}
Measurements were conducted in the HomeLab in Zwijnaarde~\cite{idlab_idlab_2024},
in two bedrooms differing in dimensions, ceiling height, and layout, with
the radar ceiling-mounted above the bed in each. Each participant
completed the activity protocol under 6 furniture configurations per
room (12 configurations total), varying target distance, aspect angle,
and clutter across sessions. The campaign included 44 participants (17 in Room~1, 24 in Room~2, 3 in
both), aged 21--86. The protocol and consent procedure were approved by
the relevant ethics committees of Ghent University, and all participants
provided written informed consent. Room~1 was released with our earlier work~\cite{lambrecht2026comparison}, and Room~2 is new to this work. Participants followed \ac{tts} instructions
describing bedroom scenarios rather than explicit activity commands, to
encourage natural transitions between actions. Each instruction was
mapped to one of eight activity classes: \emph{no activity}, \emph{walk},
\emph{stand up}, \emph{sit down}, \emph{lie on bed}, \emph{get up bed},
\emph{anxious}, and \emph{hand movement}.

\subsection{Hardware}
\label{sec:hardware}

\ac{ir-uwb} data was collected with DW3000 transceivers on Qorvo QM33120WDK1 \acp{dk}, using TX--RX pairs in a pseudo-monostatic configuration with nodes 30~cm apart. A channel~5 pair recorded in both rooms. In Room~2, an identical channel~9 pair next to it recorded the same activity executions, without interference as the bands do not overlap. Transmitter and receiver run on independent clocks, giving rise to a time-varying phase offset~\cite{demoerloose2026mmlevelaccurateuwbradar}. Packets use a 512-symbol preamble at a 64~MHz pulse repetition frequency, and each yields a \ac{cir} of $N_f = 50$ range bins spaced 1.0016~ns apart, with the \ac{los} path at bin~3. \acp{cir} were sampled at the maximal rate of 150~Hz.


\section{Methodology}
\label{sec:methodology}

The representations are compared through a common pipeline of preprocessing, windowing, classification, and grouped cross-validation. Each representation is tuned independently, since a comparison at an operating point that suits one representation measures the tuning rather than the representation.

\subsection{Evaluation Protocol}
\label{sec:eval:splits}

All models are evaluated with grouped cross-validation, in which recordings are partitioned by environment configuration so that no configuration appears in both training and test data. Two split schemes represent different deployment scenarios. \Ac{lobpo} models furniture
changes after installation, with the room and radar mounting fixed. Each of its six folds holds out two configurations from one room, three folds
per room. Since many activities occur in bed, the two held-out configurations share a similar bed position, and no training configuration
places the bed at the same position relative to the radar. \Ac{loro} models deployment in a new room. Each of its two folds holds out one room,
changing the geometry, clutter, and radar-to-bed positioning at once. \Ac{lobpo} provides enough folds for paired statistical tests and is
therefore used for all configuration selection (\cref{sec:eval:search}). \Ac{loro} has only two folds and is reported descriptively. Subject shift is not evaluated separately, as subject-independent radar has been studied extensively, whereas environmental generalization remains a more pressing challenge for practical deployment~\cite{crossenv_robustness, lambrecht2026comparison}. A leave-subjects-out axis would also multiply the cross-validation runs across all representations and configurations.

Performance is measured with macro-F1, which weights all eight classes equally despite their imbalance, computed per fold and reported as mean
and standard deviation across folds, with each fold weighted equally.

\subsection{Classifier and Fusion}
\label{sec:eval:model}

Each representation is divided into 2\,s windows with a 0.2\,s stride,
long enough to cover the posture transitions in the dataset. A
window is labeled with an activity if that activity covers at least 80\%
of it and is otherwise discarded, in both training and test, so reported
scores exclude ambiguous transitions that a deployed system would still
need to classify.

\begin{figure}[!t]
  \centering
  \includegraphics[width=\columnwidth]{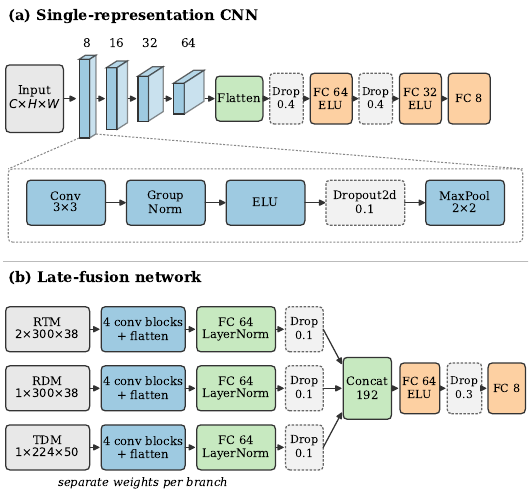}
  \caption{Classifier architectures. (a) Single-representation \ac{cnn},
used identically for each representation. The inset expands one
convolutional block. (b) Full fusion network, combining all three
representations. Each branch uses the convolutional blocks of (a) with
separate weights, and pairwise fusion uses two branches with the same
head.}
  \label{fig:models}
\end{figure}

All representations are classified with the same compact \ac{cnn} (\cref{fig:models}a). \acp{cnn} are the most common classifier in radar
\ac{har}~\cite{signal_adapted_CNN,Ayaz2025,time_freq_comparison}, and since this study compares inputs rather than models, a single shared
architecture keeps the representation as the main variable. Only the first convolutional and fully connected layers adapt to the input shape, and each model has fewer than 200k parameters. All models are trained for 50 epochs with AdamW (weight decay 0.01), batch size 64, and weighted sampling against class imbalance. The learning rate is selected per representation (\cref{sec:eval:search}) and halved when the validation loss stalls for five epochs. The checkpoint with the lowest validation loss on one held-out training configuration is evaluated.

The fusion network (\cref{fig:models}b) tests whether complementarity
between representations can be exploited for higher accuracy. Each
representation gets its own branch with independent weights and its
selected configuration, so inputs with different axes and shapes are
combined at the embedding level. Layer normalization of each embedding
prevents differences in scale from dominating the fusion.

\subsection{Configuration Search}
\label{sec:eval:search}

To compare the representations fairly, the configuration search selects the processing of each one separately, since a shared setting favors whichever representation it happens to suit. The search follows three steps, matching \cref{sec:results}. First, the processing
of each representation is selected independently. Second, the slow-time
sampling rate, Doppler bandwidth, and carrier frequency are varied one at a
time from these configurations, to measure how much each representation
tolerates different deployment settings. Third, the final configurations
are compared under layout and room shift, and fused. \Cref{tab:search}
lists all axes and candidate values.

Every setting is evaluated with full cross-validation and three learning
rates, so a single setting already requires 18 training runs on
\ac{lobpo}, and crossing all axes would quickly exceed any feasible
budget. Selection is therefore staged, varying one axis at a time with
earlier choices fixed, so the cost grows with the sum of candidates
rather than their product. Each stage is decided on \ac{lobpo}, which has
enough folds for paired statistical tests. A candidate is dropped only if it is
significantly worse than the best (\cref{sec:eval:compare}). Near-ties
stay alive into the next stage, so no early choice is fixed on a margin that a later axis could reverse. Only after the last stage is the configuration with the highest mean macro-F1 selected. Staged gridsearch can miss interactions between axes, so stages are ordered from largest to
smallest expected effect, giving early choices the widest margins, and
the learning rate is crossed with every setting. Axes with a plausible
interaction, such as clutter removal with \ac{fft}
window, or sampling rate with Doppler bandwidth, were also checked
jointly, and the conclusions did not change. \Ac{loro} is trained for the
clutter removal settings and final configurations but never informs a
decision. In total, over 1{,}000 cross-validated models were trained,
about 100 of which appear in \cref{sec:results}.

Clutter removal targets the static term $s_k$ in \cref{eq:decomposition},
the dominant indoor interference. As its variants are countless, one
estimator per family is compared with no removal: a window statistic
(mean), an empty-room reference (tail), temporal differencing
(\ac{mti}), and a post-Doppler estimate (Doppler median). The rectangular
\ac{fft} window is compared with Hann, the common
default~\cite{swinT,time_freq_comparison,Ayaz2025}, which suppresses
spectral leakage at the cost of Doppler resolution. A lower sampling rate reduces radio activity, data rate, and
computation, and Doppler cropping reduces the input size. Channels~5
and~9 differ in path loss. Channel~9 was recorded alongside
channel~5 only in Room~2, so the channel comparison is restricted to
the three \ac{lobpo} folds of that room. Three folds give too little
power for a meaningful test, so this comparison is reported
descriptively.

The \ac{tdm} \ac{cpi} length and stride are not discussed in the results (\cref{sec:results}), as their candidates did not differ significantly in macro-F1. They are fixed at the candidate with the highest mean macro-F1 in most comparisons, a \ac{cpi} of 224 pulses and a stride of 7 pulses.

\begin{table}[!t]
\centering
\caption{Configuration search space.}
\label{tab:search}
\footnotesize
\setlength{\tabcolsep}{4pt}
\renewcommand{\arraystretch}{1.3}
\begin{tabular}{|l|l|>{\raggedright\arraybackslash}p{0.40\columnwidth}|}
\hline
\textbf{Axis} & \textbf{Applies to} & \textbf{Candidate values} \\
\hline\hline
Learning rate       & all                & $3\times10^{-4}$, $10^{-3}$, $3\times10^{-3}$ \\
\hline\hline
Clutter removal     & all                & none, mean, tail, MTI,
                                           Doppler median$^{\dagger}$ \\
\hline
FFT window          & \ac{rdm}, \ac{tdm} & rectangular, Hann \\
\hline
\ac{cpi} length     & \ac{tdm}           & 128, 160, 192, 224, 256 pulses \\
\hline
\ac{cpi} stride     & \ac{tdm}           & 7, 16 pulses \\
\hline\hline
Slow-time rate      & all                & 50, 75, 100, 125, 150\,Hz \\
\hline
Doppler bandwidth   & all                & $\pm$30, $\pm$45, $\pm$60, $\pm$75\,Hz$^{\ddagger}$ \\
\hline
Channel             & all                & 5, 9 (Room~2 only) \\
\hline
\end{tabular}

\vspace{3pt}
\begin{minipage}{\columnwidth}
\footnotesize
The learning rate is crossed with every setting. The middle block is
selected in stages, in the listed order. The bottom block is varied one
at a time from the selected configuration.
$^{\dagger}$Not defined for \ac{rtm}.
$^{\ddagger}$Doppler cropping for \ac{rdm} and \ac{tdm}, slow-time low-pass
filtering for \ac{rtm}.
\end{minipage}
\end{table}
\subsection{Statistical Testing}
\label{sec:eval:compare}

Fold difficulty varies strongly. A hard fold lowers the score of every
configuration and inflates the standard deviation across folds, so two
configurations can overlap in mean $\pm$ standard deviation even when
one wins on every fold. Since all configurations share the same fold
partition, they are compared through within-fold differences in
macro-F1, in which fold difficulty cancels. The standard deviations in
the tables therefore describe the spread across folds and are not used
for comparison.

Differences on \ac{lobpo} are tested using the corrected resampled
$t$-test~\cite{nadeau_inference_2003,bouckaert_replicability_2004}. Since folds share most training
data, their scores are not independent, and a standard paired $t$-test
underestimates the variance. The corrected test uses
\begin{equation}
  \mathrm{SE} = \sqrt{\left(\frac{1}{K} +
  \overline{\frac{n_\text{test}}{n_\text{train}}}\right)\hat{\sigma}_d^{2}},
\end{equation}
where $\hat{\sigma}_d^{2}$ is the sample variance of the $K=6$ fold
differences and $\overline{n_\text{test}/n_\text{train}}$ is the average
test-to-training window ratio. The mean difference divided by
$\mathrm{SE}$ is compared with a $t$ distribution with $K-1$ degrees of
freedom. Tests are two-sided with $\alpha=0.05$, and the figures' 95\%
confidence intervals use the same standard error. Each test compares a
candidate against a reference, such as 150\,Hz for the
sampling rate, without multiplicity correction. With only two \ac{loro}
folds and three folds in the channel comparison, these results are
reported descriptively rather than statistically.

\subsection{Computational Cost}
\label{sec:eval:cost}

Computational cost is measured at two operating points per representation, both derived from the configuration selected in \cref{sec:results}. The accuracy-focused configuration prioritizes macro-F1 and uses the full 150~Hz sampling rate and Doppler band. Since some deployments prioritize latency over a small loss in accuracy, the speed-focused configuration uses the lowest sampling rate, and for \ac{rdm} and \ac{tdm} the narrowest Doppler band, whose accuracy remains statistically tied with the accuracy focused version. Fusion is timed with the focus on accuracy.

Latency is measured on an Intel Xeon Silver 4108 \ac{cpu} (1.80~GHz) with PyTorch~2.5.1 in FP32, one \ac{cpu} thread and batch size~1. Each trial comprises 600 measured inferences: ten windows from each of two recordings, with 30 repetitions after five warm-up iterations per window. Each timing is the median of ten trial-level means. A timed run covers preprocessing from the phase-referenced \acp{cir} to the model input, including clutter removal and input normalization, followed by a network forward pass. Model cost is also reported in hardware-independent \acp{macc}.

Consecutive windows overlap by 90\%, so each new window adds 0.2\,s of \acp{cir}. The \ac{rtm} needs no transform, so its per-window cost consists only of clutter removal and input normalization. The \ac{rdm} is recomputed per window, since its slow-time \ac{fft} spans the whole window. The \ac{tdm} exploits the overlap by computing only the Doppler vectors of new \acp{cpi}. Doppler-median subtraction reuses the \acp{rdm} or \ac{tdm} Doppler vectors already computed within its interval and adds only the median and subtraction. Computing the tail-subtraction background profile and initializing the \ac{cpi} cache are one-time costs and are excluded.

\begin{figure*}[!t]
  \centering
  \includegraphics[width=0.9\textwidth]{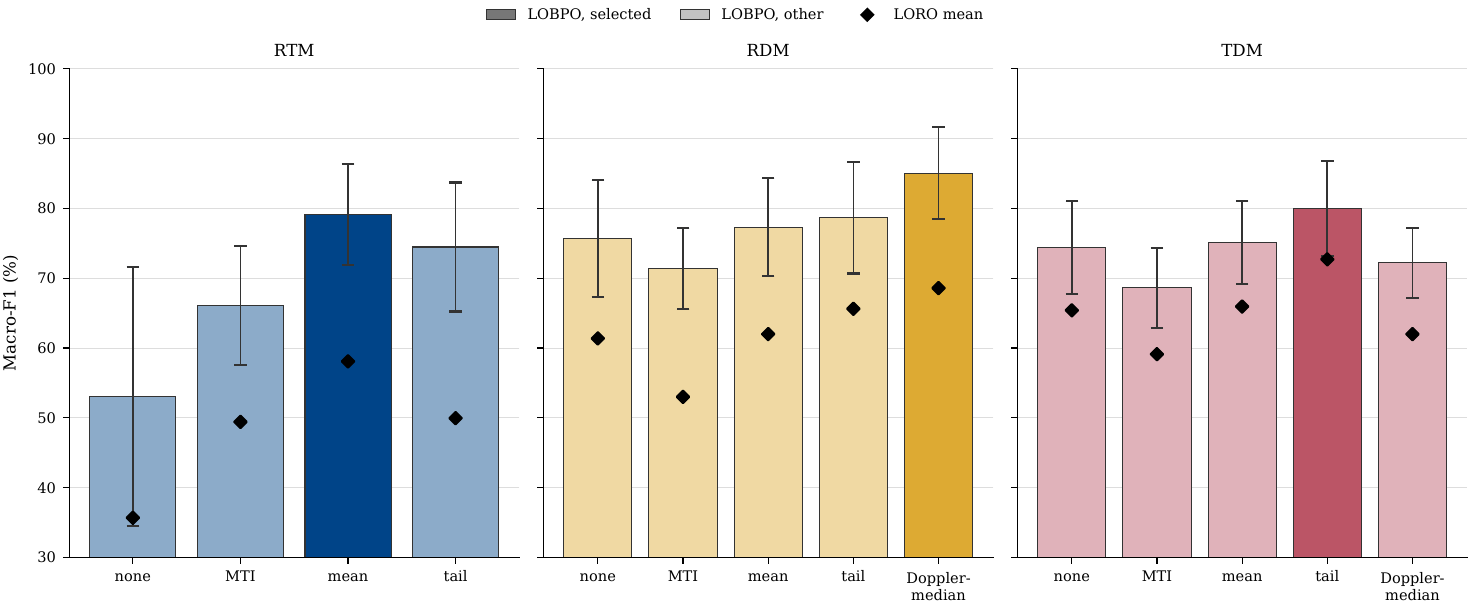}
  \caption{Macro-F1 per clutter removal method. Bars show the \ac{lobpo} mean and standard deviation across folds, with the selected method highlighted, and diamonds show the \ac{loro} mean. Each representation favors a different method, and the \ac{rtm} is most sensitive, gaining 26 points over no removal. The ranking within each representation is preserved under room shift.}
  \label{fig:clutter_removal}
\end{figure*}

\section{Results}
\label{sec:results}

\subsection{Clutter Removal}
\label{sec:results_clutter_removal}

Clutter removal is one of the most impactful preprocessing steps across all three representations, but its effect depends strongly on the representation and the removal method (\Cref{fig:clutter_removal}). \ac{rtm} benefits most, with an improvement of $+26.0$ macro-F1 points over no removal ($p=0.026$), compared with $+9.4$ for \ac{rdm} ($p<0.001$) and $+5.6$ for \ac{tdm} ($p=0.028$). This larger gain for \ac{rtm} likely stems from its more distributed clutter. While \ac{rdm} and \ac{tdm} concentrate static reflections around zero Doppler, \acp{rtm} spread them across the entire slow-time and range map, causing more overlap with the body response. The selected best methods are mean subtraction for \ac{rtm} ($+4.6$ over tail subtraction, $p=0.052$), Doppler-median subtraction for \ac{rdm} ($+6.4$ over tail subtraction, $p=0.003$), and tail subtraction for \ac{tdm}  ($+4.8$ over mean subtraction, $p=0.027$). Doppler-median subtraction, optimal for \ac{rdm}, does not improve \ac{tdm} , likely because the range collapse mixes body and clutter energy in the same Doppler bins, so the median removes more body signal. \ac{mti} does not improve standard \ac{rdm} or \ac{tdm}  and significantly lowers \ac{tdm}  performance by $5.8$ points ($p=0.022$), likely because its high-pass character attenuates the low-Doppler components of body motion along with the static clutter.

Under room shift (\ac{loro}), macro-F1 drops for all configurations, but the ranking of removal methods is preserved for each representation, with gains of $+22.4$, $+7.2$, and $+7.3$ points over non-removal for \ac{rtm}, \ac{rdm}, and \ac{tdm}, respectively. For \ac{rtm}, mean subtraction exceeds tail subtraction by $8.1$ points, supporting its selection despite the non-significant \ac{lobpo} difference. For \ac{tdm}, tail subtraction gains $6.8$ points over mean subtraction under
\ac{loro}, compared with $4.8$ under \ac{lobpo}, showing that a short empty-room reference, captured once at installation, is particularly effective when
the room changes. \ac{rdm} reaches its best room-shift result without any reference, with tail subtraction trailing Doppler-median subtraction by
$3.0$ points.


\subsection{Slow-time Sampling and Doppler Bandwidth}
\begin{figure}[!t]
  \centering
  \includegraphics[width=0.8\columnwidth]{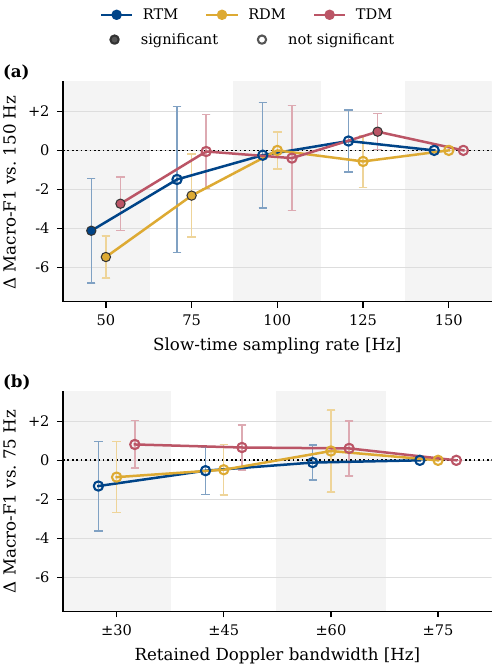}
  \caption{Change in macro-F1 relative to the reference configuration when (a) reducing the slow-time sampling rate from 150~Hz and (b) reducing the retained Doppler bandwidth from $\pm$75~Hz at a 150~Hz sampling rate. Markers show the mean paired difference across the 6 LOBPO folds with 95\% confidence intervals. Filled markers indicate a significant difference from the reference ($\alpha=0.05$).}
  \label{fig:sampling_rate}
\end{figure}

The slow-time sampling rate can be reduced substantially below the 150~Hz reference without loss, but the tolerable reduction depends on the representation (\Cref{fig:sampling_rate}a) and the activities themselves. All three representations degrade at 50~Hz, by $-4.1$ macro-F1 points for \ac{rtm} ($p=0.011$), $-5.5$ for \ac{rdm} ($p<0.001$), and $-2.7$ for \ac{tdm} ($p=0.004$). \ac{rtm} and \ac{tdm} are statistically tied with the reference from 75~Hz onward, whereas \ac{rdm} still loses $2.3$ points at 75~Hz ($p=0.038$) and only saturates at 100~Hz. Above saturation the curves are flat and non-monotonic, with all deviations below 1 point, the largest being \ac{tdm} at 125~Hz ($+1.0$, $p=0.045$). The loss at low rates is concentrated in the fast vertical transitions. Averaged over the three representations, stand up and sit down gain $+8.2$ and $+6.5$ F1 points between 50 and 150~Hz, against $+2.9$ for walk and $+1.3$ for anxious movement. Under the ceiling-mounted geometry, vertical motion is aligned with the line of sight and therefore produces the highest radial velocities in the dataset, so these transitions are likely the first to alias as the rate is lowered.

The retained Doppler bandwidth can be reduced to $\pm$30~Hz without measurable accuracy loss for any representation (\Cref{fig:sampling_rate}b). The changes are $-1.3$ points for \ac{rtm} ($p=0.200$), $-0.9$ for \ac{rdm} ($p=0.279$), and $+0.8$ for \ac{tdm} ($p=0.147$) relative to the full $\pm$75~Hz band. The classifier therefore requires only radial velocities below 0.69~m/s on channel~5, while the full band extends to higher velocities. The accuracy loss at lower sampling rates is thus unlikely to result from missing high-velocity content and is more likely caused by aliasing into the relevant Doppler range. Although cropping brings no accuracy gain, it reduces the Doppler dimension of \ac{rdm} and TDM by 60\%, which can lead to significant gains in inference time (\Cref{sec:cost}). For \ac{rtm}, the equivalent operation requires slow-time low-pass filtering but does not reduce the input size.

The rectangular window outperforms the Hann window for the \ac{rdm}, by $+7.7$ points without clutter removal ($p=0.010$) and $+4.5$ points with Doppler-median subtraction ($p=0.002$). For the \ac{tdm}, the gains are smaller and not significant ($+2.0$, $p=0.184$, and $+0.5$, $p=0.305$). Doppler resolution therefore matters more for activity recognition than leakage suppression. The Hann window widens the main lobe and merges closely spaced Doppler components, which the \ac{rdm} keeps separate per range bin. The \ac{tdm} already averages over range, so the lost separation costs less. The rectangular window is retained for both.

\subsection{Carrier Frequency}
\begin{table}[t]
\centering
\caption{Carrier frequency comparison, second room only. Positive values
favour channel 5, which is ahead in all nine pairs.}
\label{tab:channel}
\begin{tabular}{|l|c|c|c|c|}
\hline
\multicolumn{5}{|c|}{\textbf{$\Delta$ macro-F1 (ch5 $-$ ch9) [\%]}} \\
\hline
\textbf{Input} & \textbf{Fold 1} & \textbf{Fold 2} & \textbf{Fold 3} & \textbf{Mean} \\
\hline
RTM & $+1.5$ & $+10.2$ & $+2.2$ & $+4.7$ \\
RDM & $+1.1$ & $+3.9$ & $+0.6$ & $+1.9$ \\
TDM & $+9.1$ & $+16.2$ & $+8.7$ & $+11.3$ \\
\hline
\end{tabular}
\end{table}


On the three \ac{lobpo} folds of Room~2, channel~5 outperforms channel~9 in all nine
representation-fold pairs (\Cref{tab:channel}), with a mean gain of 1.9 macro-F1 points
for \ac{rdm}, 4.7 for \ac{rtm}, and 11.3 for \ac{tdm}. This is consistent with the higher
path loss and attenuation at the higher carrier, which weaken the body echo. Since all
configurations were selected on channel~5, the comparison was repeated without clutter
removal. Averaged over all three folds, the gap then widens to 11.7 points for \ac{rdm}
and 18.5 for \ac{rtm}, as static clutter masks the weaker channel~9 echo more strongly.
The advantage of channel~5 is therefore not an artifact of tuning, and channel~5 is used
throughout.
\subsection{Representation Performance Across Generalization Axes}
\label{sec}

\begin{table*}[!t]
\centering
\caption{Final per-representation configurations and macro-F1 (\%) across generalization axes. All configurations use a slow-time sampling rate of 150 Hz, the full retained Doppler bandwidth of $\pm$ 75 Hz and a rectangular window. Values are mean $\pm$ standard deviation over folds.}
\label{tab:axis_comparison}
\begin{tabular}{|l|l|l|l|l|c|}
\hline
\textbf{Representation} & \textbf{Complex input} & \textbf{Clutter estimator} & \textbf{LOBPO} & \textbf{\ac{loro}} & \textbf{$\Delta$ LOBPO $\rightarrow$ \ac{loro}} \\
\hline
RTM  & I/Q             & Mean           & $79.1 \pm 7.3$ & $58.1 \pm 16.3$ & $-21.0$ \\
\hline
RDM  & Magnitude       & Doppler median & $\mathbf{85.1 \pm 6.6}$ & $68.6 \pm 3.6$ & $-16.5$ \\
\hline
TDM & Magnitude       & Tail           & $80.0 \pm 6.8$ & $\mathbf{72.7 \pm 7.0}$ & $\mathbf{-7.3}$ \\
\hline
\end{tabular}
\end{table*}

Table~\ref{tab:axis_comparison} compares the representations under layout and room shift using their respective optimal configurations. Performance differs across the two generalization axes, with the relative ordering changing between them.

Under layout shift, \ac{rdm} achieves the highest macro-F1 at $85.1\%$, exceeding RTM by $6$ points ($p=0.031$) and TDM by $5.1$ points ($p=0.039$). RTM and TDM do not differ significantly, with a gap of $0.9$ points ($p=0.173$). Thus, under a fixed room geometry, \ac{rdm} provides the strongest generalization across layouts.

Under room shift, TDM achieves the highest macro-F1 at $72.71\%$, followed by \ac{rdm} at $68.6\%$ and RTM at $58.1\%$. The room-shift results contain only two folds and are therefore reported descriptively. The performance decrease from layout to room shift is consistent across all representations, by $7.27$ points for TDM, $16.5$ points for \ac{rdm}, and $21.0$ points for RTM. The smaller decrease for TDM indicates greater robustness to room changes, whereas RTM shows the largest degradation. This robustness does not depend
on the empty-room reference, as TDM also shows the smallest decrease with
reference-free mean subtraction ($9.2$ points). This ordering is consistent with the amount of room-specific spatial structure retained by each representation, since TDM aggregates across range while RTM and \ac{rdm} preserve range structure.

\subsection{Representation Complementarity and Fusion}
\begin{table}[t]
\centering
\caption{Fusion ablation results in macro-F1 (\%). Mean foldwise difference ($\Delta$) relative to the best single representation for each evaluation axis (underlined). * indicates a statistically significant difference from the corresponding best single representation ($p<0.05$) on LOBPO; \ac{loro} differences are descriptive.}
\label{tab:fusion}
\footnotesize
\setlength{\tabcolsep}{3pt}
\begin{tabular}{|l|c|c|c|c|c|}
\hline
\textbf{Input} & \textbf{Params (k)} & \textbf{LOBPO} & \textbf{$\Delta$} & \textbf{\ac{loro}} & \textbf{$\Delta$} \\
\hline\hline
RTM  & 174.6 & 79.1 $\pm$ 7.3 & $-$5.9* & 58.1 $\pm$ 23.1 & $-$14.6 \\
\hline
RDM  & 174.5 & 85.1 $\pm$ 6.6 & \underline{ref} & 68.6 $\pm$ 5.1 & $-$4.1 \\
\hline
TDM & 199.1 & 80.0 $\pm$ 6.8 & $-$5.1* & 72.7 $\pm$ 9.9 & \underline{ref} \\
\hline\hline
RTM+RDM  & 353.4 & 85.7 $\pm$ 6.0 & +0.6 & 68.4 $\pm$ 4.8 & $-$4.3 \\
\hline
RTM+TDM & 378.0 & 85.6 $\pm$ 6.6 & +0.6 & 73.7 $\pm$ 13.3 & +1.0 \\
\hline
RDM+TDM & 377.9 & 88.5 $\pm$ 6.6 & +3.4* & 78.7 $\pm$ 8.5 & +6.0 \\
\hline\hline
Full-Fusion & 554.3 & \textbf{89.1 $\pm$ 6.9} & \textbf{+4.0*} & \textbf{80.0 $\pm$ 8.7} & \textbf{+7.2} \\
\hline
RDM$\times$3 & 529.7 & 82.4 $\pm$ 6.4 & $-$2.6* & 66.6 $\pm$ 7.8 & $-$6.2 \\
\hline
\end{tabular}
\end{table}

\begin{figure}[!t]
  \centering
  \includegraphics[width=01\columnwidth]{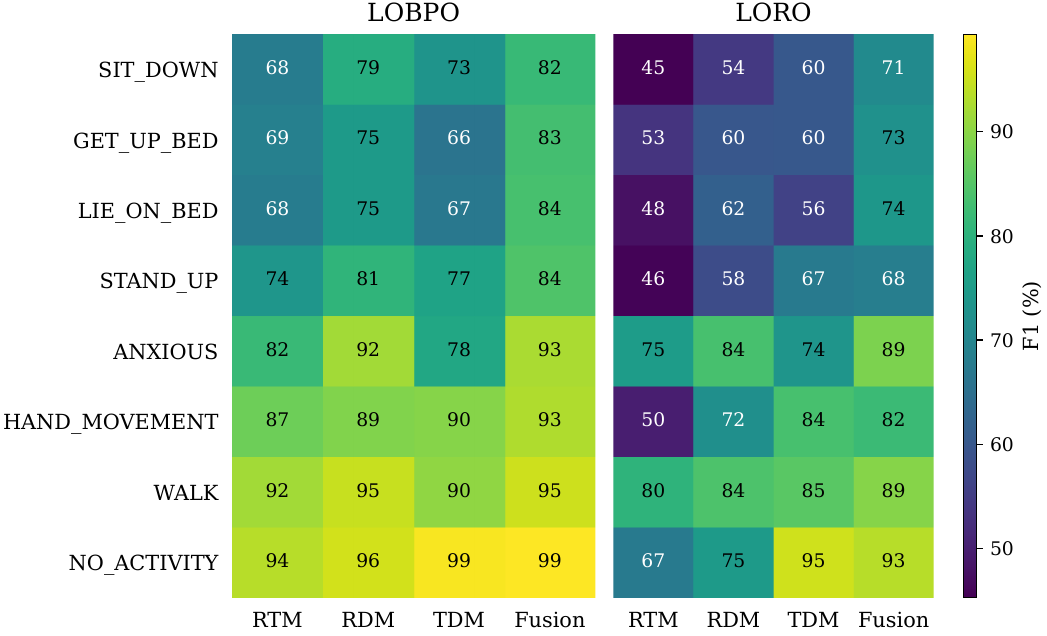}
  \caption{RDM and TDM show complementary strengths across activities, which fusion exploits to improve performance across evaluation settings. Per-activity fold-averaged F1 scores (\%) for individual and fused representations under LOBPO and \ac{loro}.}
  \label{fig:fusion_activity}
\end{figure}

Fusion improves macro-F1 across both generalization axes (\Cref{tab:fusion}). Full fusion of RTM, RDM, and TDM reaches 89.1 under LOBPO, 4.0 points above RDM, the best single representation ($p=0.003$). Under \ac{loro}, it reaches 80.0, 7.2 points above TDM. Most of this gain comes from RDM+TDM, which reaches 88.5 under LOBPO ($+3.4$, $p=0.002$) and 78.7 under \ac{loro}, within 0.6 and 1.3 points of full fusion. RTM adds little once RDM is present. RTM+RDM stays within 0.6 points of RDM on both axes, and adding RTM to RDM+TDM gains only 0.6 and 1.3 points. The fusion gain is not explained by parameter count alone. A matched-capacity model (RDM$\times$3), with three branches processing the same RDM input, scores below the best single representation on both axes, by 2.6 points under LOBPO ($p=0.020$) and 6.2 under \ac{loro}. This suggests that adding a complementary representation is more effective than simply increasing the number of branches of the same representation.

The per-activity results confirm the complementarity of RDM and TDM (~\Cref{fig:fusion_activity}). RDM outperforms RTM on every activity under both axes, so RTM contributes little that RDM does not already capture. RDM and TDM instead lead on different activities. Under \ac{loro}, TDM performs better for sit down (60 versus 54), stand up (67 versus 58) and hand movement (84 versus 72), likely because aggregating over range reduces sensitivity to the person's range distribution, which changes between rooms. RDM performs better for anxious movements under both axes (92 versus 78 under LOBPO and 84 versus 74 under \ac{loro}). During anxious movement, body parts (e.g. legs, arms, head) likely move at different ranges with distinct Doppler patterns, which RDM preserves and TDM aggregates.

The greatest full-fusion gains occur for posture transitions. Get up from bed and lie on bed improve by 8 and 9 points over the best single representation under LOBPO, and by 13 and 12 points under \ac{loro}, while sit down gains 11 points under \ac{loro}. These transitions likely combine a change in body position across range, which RDM resolves, with a motion evolution that TDM captures robustly across rooms. Walk gains at most 4 points, as all representations already recognize it well. Where TDM already leads, such as with hand movement and the no-activity class under \ac{loro}, full fusion remains within 2 points, indicating limited room for improvement.

\subsection{Representation Computing and Inference Cost}
\label{sec:cost}
At the accuracy focus, the networks take 5.6 to 7.0~ms for all three representations, so latency differences mainly stem from preprocessing (\Cref{tab:representation_complexity_final}). The \ac{rtm} needs no transform and spends only 0.37~ms on preprocessing. The Doppler-median \ac{rdm} is the slowest representation, at 14.86~ms, with more than half spent on the element-wise median over the \acp{rdm} in the 10~s interval (\cref{eq:doppler_median}). Tail subtraction cuts this preprocessing to 0.88~ms and halves the total latency to 7.56~ms, but costs 6.4 macro-F1 points (\cref{fig:clutter_removal}). Recomputing every \ac{cpi} per window would make the \ac{tdm} even slower, at 23.17~ms. Reusing the Doppler vectors of overlapping windows instead reduces its preprocessing 8.4-fold to 2.08~ms and its total latency to 7.65~ms, on par with the \ac{rtm}.

Reducing the input size accelerates all three representations at little cost. The speed-focused configurations lower latency by 38\%, 51\% and 43\% for the \ac{rtm}, Doppler-median \ac{rdm} and \ac{tdm}, losing at most 1.5 macro-F1 points. The \ac{rtm} gains least, as only its sampling rate can be reduced without adding a filter (\cref{sec:preprocessing}). For the \ac{rdm}, reducing the input is also more effective than changing the clutter removal. The speed-focused Doppler-median \ac{rdm} is faster than the accuracy-focused tail-subtraction \ac{rdm} (7.23 versus 7.56~ms) and 5.3 points more accurate. Combining both gives the lowest latency, 4.30~ms at 78.1 macro-F1, but the speed-focused \ac{tdm} is only 0.1~ms slower and 1.3 points more accurate.

Fusion reaches the highest macro-F1, 89.1, at 31.24~ms. Its inference time is close to the sum of the three single models, and the Doppler-median \ac{rdm} branch dominates its preprocessing. The 4.0-point gain over the best single representation thus costs about twice the \ac{rdm} latency. With a 0.2~s window stride, a new prediction is due every 200~ms. All configurations meet this budget, with fusion using about 16\% of it.


\begin{table}[t]
\centering
\caption{Computational cost and \ac{lobpo} macro-F1 (\%) at the
accuracy-focus (Acc.) and speed-focus (Speed) operating points. $f_s$ and
$f_{\mathrm{crop}}$ in Hz. Clutter removal: Mean, mean subtraction; DM,
Doppler-median subtraction; Tail, tail subtraction; fusion uses the
selected method of each input. Best value per column in bold.}
\label{tab:representation_complexity_final}
\footnotesize
\setlength{\tabcolsep}{2pt}
\begin{tabular}{|l|l|l|c|c|c|c|c|c|c|}
\hline
 & & & &
 & \multicolumn{3}{c|}{\textbf{Latency (ms)}} & \textbf{MMAC} & \\
\cline{6-8}
\textbf{Input} & \textbf{Set.} & \textbf{Clutter} & \boldmath$f_s$ & \boldmath$f_{\mathrm{crop}}$
 & \textbf{Prep.} & \textbf{Model} & \textbf{Total}
 & \textbf{Model} & \textbf{mF1} \\
\hline\hline
\multirow{2}{*}{RTM}
 & Acc.  & \multirow{2}{*}{Mean} & 150 & --      & 0.37 & 7.02 & 7.40 & 10.9 & 79.1 \\
\cline{2-2}\cline{4-10}
 & Speed &                       & 75  & --      & \textbf{0.29} & 4.29 & 4.58 & 5.4 & 77.6 \\
\hline\hline
\multirow{4}{*}{RDM}
 & Acc.  & \multirow{2}{*}{DM}   & 150 & --      & 8.14 & 6.72 & 14.86 & 10.1 & 85.1 \\
\cline{2-2}\cline{4-10}
 & Speed &                       & 100 & $\pm$30 & 3.56 & 3.68 & 7.23  & 4.1  & 83.9 \\
\cline{2-10}
 & Acc.  & \multirow{2}{*}{Tail} & 150 & --      & 0.88 & 6.69 & 7.56  & 10.1 & 78.6 \\
\cline{2-2}\cline{4-10}
 & Speed &                       & 100 & $\pm$30 & 0.70 & 3.60 & \textbf{4.30} & 4.1 & 78.1 \\
\hline\hline
\multirow{2}{*}{TDM}
 & Acc.  & \multirow{2}{*}{Tail} & 150 & --      & 2.08 & 5.57 & 7.65 & 8.7  & 80.0 \\
\cline{2-2}\cline{4-10}
 & Speed &                       & 75  & $\pm$30 & 1.47 & \textbf{2.92} & 4.40 & \textbf{2.9} & 79.4 \\
\hline\hline
Fusion & Acc. & -- & 150 & -- & 11.29 & 19.95 & 31.24 & 31.4 & \textbf{89.1} \\
\hline
\end{tabular}
\end{table}

\section{Discussion and Limitations}
\label{sec:discussion}

Range information is both the strongest cue and the main liability of an
\ac{ir-uwb} representation. Within a known room, the range axis separates
body parts and positions, favoring the \ac{rdm}. In a new room, the same
range structure likely becomes room-specific, and the \ac{tdm}, which
averages it out, degrades the least. The best representation is
therefore a deployment decision, not a fixed property. The \ac{tdm}, widely
used in radar \ac{har}~\cite{complex_CNN,Ayaz2025}, is the better default
for unseen rooms but sacrifices 5 macro-F1 points within a known room. This supports the choice of El Hail et al.~\cite{el_hail_radar-based_2025} for the \ac{tdm} in cross-room settings.
The \ac{rtm}, previously found or selected as the best input in
\ac{ir-uwb} studies~\cite{complex_CNN,swinT}, is here the weakest and
least robust, possibly because it relies more on fine range detail and
effective clutter removal. The \ac{rdm} and \ac{tdm} provide complementary
information, consistent with Ding et al.~\cite{ding2021}. Their fusion
outperforms both individually, whereas a matched-capacity model does not,
indicating that the gain comes from combining range-resolved and
range-averaged motion rather than increasing model size.

These differences only emerge with per-representation tuning. Preprocessing
can shift macro-F1 by as much as the gap between representations:
clutter removal improves the \ac{rtm} by 26 points, while the Hann window,
widely used in prior work~\cite{swinT,time_freq_comparison,Ayaz2025},
reduces the \ac{rdm} score by 4.5 points, nearly eliminating its lead over
the \ac{tdm}. Comparisons using a shared pipeline~\cite{ding2021,Ayaz2025}
may therefore rank preprocessing choices rather than representations.
This sensitivity is particularly relevant for commodity hardware. The
DW3000 offers three to five times less bandwidth than the research radars
used in prior work, increasing overlap between body and clutter returns
within each range bin. Despite this limitation, it retains enough range
resolution to favor the \ac{rdm}. Moreover, much of the processing
complexity is unnecessary: reducing the slow-time sampling rate and
cropping the retained Doppler bandwidth to $\pm$30\,Hz cuts network
\acp{macc} by two- to threefold and latency by about half, at a cost of
at most 1.5 macro-F1 points. This leaves headroom for embedded deployment.

The main open problem is generalization to unseen rooms. Representation
choice narrows but does not close this gap, as even full fusion loses
9 macro-F1 points under room shift. Closing it may require training on
more diverse environments or adapting to a new room through unsupervised
domain adaptation. Our evaluation covers only two rooms with the same
ceiling mounting; additional rooms, and mounting geometries are needed to assess whether the ranking generalizes. Such data is costly
to collect: our dataset involved 44 participants across 12 configurations.
Releasing it can support broader comparisons. The model search was also
bounded by computational cost. Although over 1{,}000 cross-validated
models were trained, the study covered one estimator per clutter-removal
family and one compact \ac{cnn}. Deeper and attention-based models may
exploit range information differently. Finally, latency was measured on a
server CPU, and deployment on embedded hardware remains to be evaluated,
including fusion of the low-complexity configurations.


\section{Conclusion}
\label{sec:conclusion}

In radar \ac{har}, the input representation is usually chosen once and evaluated at a single processing setting. This paper compared \acp{rtm}, \acp{rdm}, \acp{tdm}, and their fusion on an open dataset of 44 participants in 12 furniture configurations, recorded with one ceiling-mounted IEEE 802.15.4z-compliant radar. Each representation was tuned independently over five processing factors, training more than 1{,}000 cross-validated models, and evaluated under layout and room shift together with its computational cost. Processing changed macro-F1 by up to 26 points, more than any gap between representations, so a fair comparison requires per-representation tuning. With that tuning, no single representation is best. The \ac{rdm} leads within a known room and the \ac{tdm} in an unseen one. The representations are complementary, as fusion raises macro-F1 by 4.0 and 7.2 points over the best single representation in these settings, to 89.1\% and 80.0\%, a gain that added model capacity does not provide. Low-rate, band-limited processing cuts latency by 38--51\% at a cost of at most 1.5 points. In practice, the \ac{rdm} suits deployments with labeled data from the target room, the \ac{tdm} those without, and fusion those that can afford about twice the \ac{rdm} latency.

\section*{Acknowledgment}

During the preparation of this work, Microsoft Copilot and Claude (Anthropic) were used to improve the clarity of the language and figures throughout the manuscript, and GitHub Copilot was used as a programming aid for parts of the data-processing and machine-learning implementation. All AI-assisted material was reviewed and validated by the authors, who take full responsibility for the final work.

\bibliographystyle{IEEEtran}
\bibliography{references.bib}

\begin{IEEEbiographynophoto}
{Jelle De Moerloose} received the M.Sc. degree in Information Engineering Technology from Ghent University, Belgium, in 2025. He was awarded the Boulvin-Van Engelen Prize for ranking first among 363 students across all engineering technology programs over the full bachelor’s and master’s track. He then joined the Internet Technology and Data Science Laboratory (IDLab), Ghent University–imec, where he is currently pursuing the Ph.D. degree. In 2026, he was a Visiting Student Research Collaborator at Princeton University under the EVOLVE Marie Skłodowska-Curie project. His research interests include radar-based human activity recognition using ultra-wideband (UWB) technology, representation learning, weak supervision, and foundation models for RF sensing.
\end{IEEEbiographynophoto}

\begin{IEEEbiographynophoto}
{Anton Lambrecht} received the M.Sc. degree in Computer Science Engineering from Ghent University, Belgium, in 2024. Shortly thereafter, in September 2024, he joined the Department of Information Technology (INTEC) at Ghent University as a researcher within the IDLab research group, where he is currently pursuing the Ph.D. degree. His research interests include RF-based sensing using Ultra-wideband technology, with a focus on healthcare applications through machine learning techniques.
\end{IEEEbiographynophoto}

\begin{IEEEbiographynophoto}
{Adnan Shahid} (Senior Member, IEEE) received the Ph.D. degree in Information and Communication Engineering from Sejong University, South Korea, in 2015. He is currently an Assistant Professor with the Internet Technology and Data Science Laboratory (IDLab), Ghent University–imec, where he leads the “AI/ML for Wireless” research within the IDLab Intelligent Wireless Networking (iWINe) group. He actively contributes to international standardization efforts, including the IEEE P1900.8 Working Group on Training, Testing, and Evaluating Machine-Learned Spectrum Awareness Models and the ATIS Working Group on Generative AI in Telecommunications. He has participated in several major research initiatives, including the DARPA Spectrum Collaboration Challenge (SC2) and multiple H2020, Horizon Europe, and ESA projects. He has authored over 150 publications in leading conferences and journals. His research interests include wireless physical-layer foundation models, decentralized and federated learning, radio resource management, 5G/6G networks, and non-terrestrial networks (NTN).
\end{IEEEbiographynophoto}

\begin{IEEEbiographynophoto}
{Eli De Poorter} is professor at the IDLab research group from imec \& Ghent University (https://idlab.technology). His team performs research on wireless communication technologies such as (indoor) localization solutions, connected healthcare, wireless IoT solutions and machine learning for wireless systems. He performs both fundamental and applied research. He currently has around 350 publications and has a h-index of 50 (google scholar) with around 1400 citations per year. Since 2021, Prof. De Poorter has been included yearly in the Stanford University (Stanford, CA, USA), World’s Top 2\% Scientists in the domain of Networking \& Telecommunications.
\end{IEEEbiographynophoto}

\EOD

\end{document}